\documentclass{article}
\usepackage{spconf,amsmath,graphicx}
\usepackage{multirow}
\usepackage{booktabs}
\usepackage{subcaption}

\usepackage[textfont=it,tableposition=top]{caption}
\usepackage{color}

\title{Incremental Layer-wise Self-Supervised Learning for Efficient Speech Domain Adaptation On Device}
\name{
\parbox{.95\linewidth}{\centering
Zhouyuan Huo, Dongseong Hwang, Khe Chai Sim, Shefali Garg, Ananya Misra, Nikhil Siddhartha, Trevor Strohman, Françoise Beaufays}}
\address{Google LLC, USA}
\begin{document}
%
\maketitle
\begin{abstract}
Streaming end-to-end speech recognition models have been widely applied to mobile devices and show significant improvement in efficiency. These models are typically trained on the server using transcribed speech data. However, the server data distribution can be very different from the data distribution on user devices, which could affect the model performance. There are two main challenges for on device training, limited reliable labels and limited training memory. While self-supervised learning algorithms can mitigate the mismatch between domains using unlabeled data, they are not applicable on mobile devices directly because of the memory constraint. In this paper, we propose an incremental layer-wise self-supervised learning algorithm for efficient speech domain adaptation on mobile devices, in which only one layer is updated at a time. Extensive experimental results demonstrate that the proposed algorithm obtains a  Word Error Rate (WER) on the target domain $24.2\%$ better than supervised baseline and costs  $89.7\%$ less training memory than the end-to-end self-supervised learning algorithm. 
\end{abstract}
\section{Introduction}
\label{sec:intro}

Streaming end-to-end speech recognition models have been widely applied on mobile devices and show significant improvement in efficiency \cite{he2019streaming,li2021better}. The deployed models trained using the transcribed speech data in the server could perform badly if the speech data distribution on the user's device is very different from the training data. We can treat adapting the model from the server data distribution (source domain) to the on-device data distribution (target domain) as a domain adaptation problem, where the domain discrepancy may come from different application domains, noise conditions or utterance lengths.

The speech domain adaptation task has been widely researched, including transfer learning \cite{Bengio2012}, fine-tuning factorized hidden layer \cite{Sim2018}, or generative adversarial networks~\cite{Ganin2016}. Recent results have shown that self-supervised pre-training on untranscribed speech data improve speech recognition accuracy without much supervised data \cite{misra2021comparison}, including Autoregressive Predictive Coding (APC) \cite{Chung2020_apc},  Contrastive Predictive Coding (CPC) \cite{oord2018representation} and Wav2vec2.0~\cite{baevski2020wav2vec}. Federated learning \cite{mcmahan2017communication,kairouz2019advances} enables us to update models on mobile devices without compromising data privacy by keeping the training data decentralized. It has been researched or successfully deployed in different applications, e.g. emoji prediction \cite{ramaswamy2019federated}, next-word prediction \cite{hard2018federated}, and speech recognition \cite{guliani2021training}. There are two main challenges for federated speech domain adaptation on mobile devices: limited reliable labels and limited training memory. The former can be mitigated using self-supervised learning algorithms, which have been shown to be effective for domain adaptation using unlabeled data~\cite{misra2021comparison}. To address the memory requirement for on-device training, we propose using an incremental layer-wise training approach.

Layer-wise unsupervised pre-training followed by supervised fine-tuning enabled the successful training of deep neural networks before 2010s \cite{hinton2006fast}. Authors of \cite{erhan2010does} empirically show that the layer-wise unsupervised pre-training can serve as a regularizer and adds robustness and generalization to deep neural networks, which does not disappear with more data. Recent research on layer-wise learning focuses more on faster distributed training, including both supervised learning \cite{huo2018decoupled} and unsupervised learning \cite{lowe2019putting}.  However, utilizing the layer-wise technique to reduce training memory for efficient speech domain adaptation task on mobile devices is still an under-explored area.

In this paper, we propose an incremental layer-wise self-supervised learning algorithm to reduce the training memory for efficient speech domain adaptation on mobile devices. Specifically, we update only one or a few selected layers at each step and all the layers are pre-trained in an incremental schedule from bottom to top after given training steps. Training memory increases with input length and the benefit of self-supervised algorithms over supervised or semi-supervised training is that self-supervised losses can work with truncated inputs. Extensive experimental results demonstrate that the proposed algorithm obtains a  Word Error Rate (WER) on the target domain $24.2\%$ better than supervised baseline and costs  $89.7\%$ less training memory than the end-to-end self-supervised learning algorithm. 

The rest of the paper introduces three self-supervised losses in Section \ref{sec:related}, describes our proposed algorithm in Section \ref{sec:alg}, specifies the experimental setup (model, data and losses) in Section \ref{sec:exp}  and presents the results  in {Section ~\ref{sec:results}}. We conclude in {Section~\ref{sec:conclusion}}.

\section{Self-Supervised Losses}
\label{sec:related}
There are three popular self-supervised losses for speech, Contrastive Predictive Coding (CPC) \cite{oord2018representation},  Wav2vec2.0 \cite{baevski2020wav2vec} and Autoregressive Predictive Coding (APC) \cite{Chung2020_apc},. 

CPC \cite{oord2018representation} minimizes a probabilistic contrastive loss to induce the latent space to capture information that is maximally useful to discriminate future samples from negative samples. In the paper, we let CPC predict the future 12 frames and discriminate each future frame from $8$ negative frames sampled from the same sequence. 

Wav2vec2.0 \cite{baevski2020wav2vec}, on the other hand, learns a good representation to discriminate the latent vectors of masked samples from negative samples.  
Because we use stacked log-Mel spectrograms as the input, we do not add convolutional layers to extract features as in the original paper. We also remove the quantization module and use the log-Mel feature as the latent vector since no differences are observed in our task. 

Unlike the contrastive loss in CPC or Wav2vec2.0, APC  \cite{Chung2020_apc} is trained to encode the given speech utterances and predict future frames, where l2 or l1 loss is applied. In the paper, we use l2 loss and add a total variation regularization \cite{rudin1994total} between consecutive frames, with regularization weight 0.1.

\section{Incremental layer-wise self-supervised learning}
\label{sec:alg}

One major issue with federated learning of speech models is the limited memory and compute resources available on mobile devices. Self-supervised learning offers the flexibility of attaching the loss to an arbitrary layer of the encoder and truncating the input sequence since there is no need to worry about the alignment between the input and label sequences (which is needed for supervised training). This flexibility allows the training memory to be controlled accordingly. 

Self-supervised learning can be performed in a layer-wise manner, where one or more selected encoder layers are updated at each step. By attaching the self-supervised loss on top of the selected encoder layers, the layers above the selected layers need not be included in the training graph, and there is no need to compute the gradients for the layers below the selected layers. To reduce the training memory consumption, we propose the incremental layer-wise self-supervised learning algorithm, which updates only one or a few selected layers at each step and all the layers are pre-trained in an incremental schedule from bottom to top where a block of layers is pre-trained for a predefined number of steps before moving on to the next block.

\begin{figure}[t]
 \centering
\includegraphics[width=0.38\textwidth]{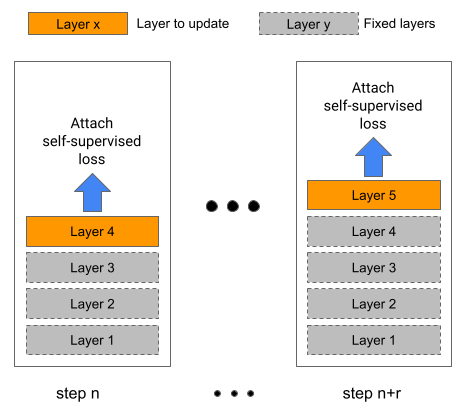}
\caption{Incremental layer-wise self-supervised learning.}
\label{fig:alg}
\vspace{-0.4cm}
\end{figure}

Figure \ref{fig:alg} visualizes the proposed incremental layer-wise self-supervised learning algorithm. At step $n$, we fix the parameters of the first three layers, drop the layers after layer 4 and update layer 4 by attaching a self-supervised loss on top of it. Since layers 1 to 3 are not updated, we only need to compute the gradients for layer 4. After pre-training layer 4 for $r$ steps, we fix the parameters of the first four layers and update layer 5 only by attaching a self-supervised loss on top of it. All the layers can be updated from the bottom layers to the top layers progressively until all the layers are pre-trained. The incremental layer-wise self-supervised learning algorithm offers us a memory- and compute-efficient way to perform unsupervised domain adaptation, which is particularly important for on device learning where human transcriptions are not available. In the next sections, we will demonstrate that our algorithm can achieve comparable results to the end-to-end self-supervised learning with much less memory.

The incremental layer-wise self-supervised learning algorithm is very memory-efficient since we only update one layer at each step without storing activations for all the fixed layers. In the best case, the training memory requirement for updating layer $1$ is roughly $\frac{1}{L}$ the training memory required to update all the $L$ layers in an end-to-end fashion. When updating the intermediate layer $l$, we can save time and memory by performing the feed-forward of the first $l-1$ layers using quantized weights. 

\section{Experimental Setup}
\label{sec:exp}


\subsection{Model Architecture Details}
\label{sec:models}
Following \cite{gulati2020conformer,li2021better}, we use a Conformer RNN-T model in the experiments. The encoder has $17$ Conformer layers with model dimension $512$. To restrict the  model from using any future information as \cite{li2021better}, each conformer layer  is streaming and contains causal convolution, left-context attention layers and feed-forward dense layers. The convolution kernel size is $15$ and the self-attention layer consists of $8$ heads with $65$ left context length. As the model is causal, the right context length is 0. The RNN-T decoder consists of a prediction network with $2$ LSTM layers with $2048$ units projected down to $640$ output units, and a joint network with a single feed-forward layer with $640$ units. The model is trained to predict $4,096$ word pieces.  The model input is a vector of size $528$, consisting of $4$ contiguous frames of $128$-dimension logMel features \cite{narayanan2019recognizing} sub-sampled by a factor of $3$ and one-hot domain-id vector of size $16$.

\subsection{Data Sets}
\label{sec:data}
The speech domain adaptation task in the paper utilizes the multi-domain utterances as described in \cite{narayanan2019recognizing}.
These multi-domain (MD) utterances span domains of search,
farfield, telephony and YouTube. All datasets are anonymized and
hand-transcribed. As shown in Table \ref{tab:data}, we use Medium-form ({MF}) utterances as target domain without transcriptions and the complement of MF in MD as the source domain with transcriptions.
To make the Conformer RNN-T model general to both the source and target domains, we first pre-train the encoder layers of Conformer RNN-T model on MD data using self-supervised learning, and then fine-tune the encoder and decoder layers using the RNN-T loss on the source domain only (MD-MF). For performance evaluation, we calculate the Word Error Rate (WER) of Medium-form (MF) to measure the performance on target domain and the WER of Short-form (SF) for the performance on the source domain. 

\begin{table}[t]
    \centering
    \caption{Overview of training data sets. MD-MF denotes the source domains utterances with transcriptions, MF denotes the target domain utterances without transcriptions. MD-MF and MF are subsets of MD. SF is a subset of MD-MF. }
    \label{tab:data}
    \begin{tabular}{ccc}
        \toprule
        \textbf{Data set} & \textbf{Transcribed} & \textbf{Hours} \\
        \midrule
         Multi-domain ({MD}) & Yes $\&$ No & $400$k \\
        \midrule
         Medium-form ({MF}) & No & $26$k \\
         MD - MF  & Yes & $374$k \\
         Short-form (SF)  & Yes & $27$k \\
        \bottomrule
    \end{tabular}
\end{table}

\section{Experimental results}
\label{sec:results}
In this section, we conduct extensive experiments to evaluate the incremental layer-wise self-supervised learning algorithm on the speech domain adaptation task. 

\subsection{End-to-End Self-supervised Learning}
\label{sec:exp_ssl}
In Table \ref{tab:e2e}, we compare the performance of supervised, semi-supervised and three end-to-end self-supervised learning algorithms. The first row is the baseline model trained with supervised learning on the source domain (MD-MF) only. The second row shows the results of semi-supervised learning where the teacher is the supervised model trained on MD-MF. 
For self-supervised learning, we follow the pre-training and fine-tuning schedule. We first perform self-supervised pre-training on MD data, and then fine-tune on the source domain (MD-MF) only. It is evident that self-supervised pre-training using APC, CPC, or Wav2vec2.0 algorithms can improve the WER performance on the target domain (MF) by a large margin compared to the supervised baseline using source domain data only. They also perform as well or better than the baseline on the source data. We observe that the uni-directional self-supervised objective functions (APC, CPC) works better than the bi-directional objective function (Wav2vec2) when training a streaming model.

\begin{table}[!t]
    \centering
    \caption{Comparisons of Word Error Rate (WER) between supervised, semi-supervised baselines and three self-supervised learning algorithms. In parenthesis, the relative improvement with respect to the supervised baseline.}  
    \label{tab:e2e}
    \begin{tabular}{cccc}
        \toprule
        {\textbf{Algorithms}}& \multicolumn{2}{c}{\textbf{Word Error Rate ($\%$)}} \\
         &  {\textbf{MF}} & {\textbf{SF}} \\
        \midrule
        {Supervised} & $6.2$  & {$6.3$} \\
        {Semi-Supervised } &  $4.9$ & $6.1$ \\
        \midrule
        APC + Fine-tune & $\textbf{4.5}$ ($\textbf{-27.4\%}$) & {$6.1$} \\
        CPC +  Fine-tune &  $\textbf{4.5}$ ($\textbf{-27.4\%}$) & {$6.1$} \\
        {Wav2vec2.0 +  Fine-tune}& {${4.6}$} (${-25.8\%}$) & {$6.3$} \\
        \bottomrule
    \end{tabular}
\end{table}

\subsection{Layer-wise Self-Supervised Learning}
\label{sec:exp_lw_ssl}

The results in the last section show that end-to-end self-supervised learning algorithms can improve the performance on the target domain in the speech domain adaptation task. In Table \ref{tab:exp_lw_ssl}, we compare the proposed incremental layer-wise (ILW) and greedy layer-wise (GLW)~\cite{lowe2019putting} self-supervised learning algorithms. GLW attaches self-supervised losses on top of every encoder layer and updates all the layers at every step, which requires much more memory in the training.  By contrast, IWL only updates one layer at each step and pre-train the bottom layers with more steps than top layers. For the encoder with $17$ layers, ILW updates the model from layer $1$ to layer $17$ in an incremental way, and the number of steps to update each layer is presented in Figure \ref{fig:pretrainsteps}. All the results in Table \ref{tab:exp_lw_ssl} are obtained by running self-supervised pre-training on all the multi-domain (MD) data and then fine-tuning on the source domain (MD-MF).

In Table \ref{tab:exp_lw_ssl}, we can observe that the MF WER results of GLW and ILW algorithms are comparable and both achieve $4.6\%$ WER on the target domain (MF) by using the CPC loss. APC performs the worst because applying reconstruction losses with the same targets on all the layers make them learn similar representations. The disadvantage of Wav2vec2.0 loss is the model mismatch between bi-directional self-supervised learning and uni-directional fine-tuning. In terms of training memory consumption, ILW is much more efficient than GLW because ILW only updates one layer at each step and all the activations of the fixed layers are dropped after the forward pass.
Since the CPC method obtains the best performance, we will use it to investigate the effect of hyper-parameters on incremental layer-wise CPC algorithm in the following sections.

\begin{table}[t]
    \centering
    \caption{Comparison between greedy layer-wise and incremental layer-wise scheme for three self-supervised learning algorithms. We pre-train on MD data and then fine-tune on source domain (MD-MF) only.}
    \label{tab:exp_lw_ssl}
    \begin{tabular}{cccc}
        \toprule
       \textbf{Layer-wise} &  {\textbf{Pre-training}} & \multicolumn{2}{c}{\textbf{Word Error Rate ($\%$)}} \\
        \textbf{scheme} & {\textbf{algorithm}} &  \textbf{MF} & \textbf{SF} \\
        \midrule
        \multirow{3}{*}{Greedy} & APC & {$5.0$} & {$6.3$} \\
         & CPC  & $\textbf{4.6}$ & $\textbf{6.1}$ \\
         & Wave2vec2 & {$4.8$} & {$6.2$} \\
         \midrule
        \multirow{3}{*}{Incremental} & APC & {$5.1$} & {$6.2$} \\
         & CPC  & {$\textbf{4.6}$} & {$\textbf{6.1}$} \\
         & Wave2vec2 & {$5.1$} & {$6.2$} \\
        \bottomrule
    \end{tabular}
\end{table}

\subsection{The Effect of Hyper-Parameters}
We investigate three important hyper-parameters for the incremental layer-wise CPC algorithm and reduce the training memory significantly without compromising performance.

\subsubsection{Incremental Layer-wise Pre-Training Schedules}
\label{sec:exp_schedules}

\begin{figure}[h]
     \centering
     \begin{subfigure}[b]{0.23\textwidth}
         \centering
         \includegraphics[width=\textwidth]{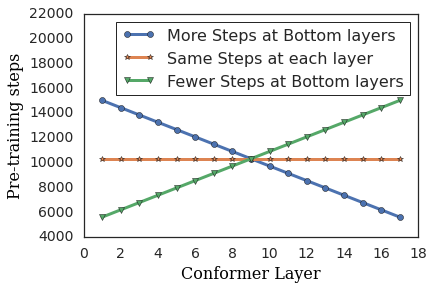}
         \caption{Figure 2a: Pre-training steps per layer for three schedules.}
         \label{fig:pretrainsteps}
     \end{subfigure}
     \begin{subfigure}[b]{0.23\textwidth}
         \centering
         \includegraphics[width=\textwidth]{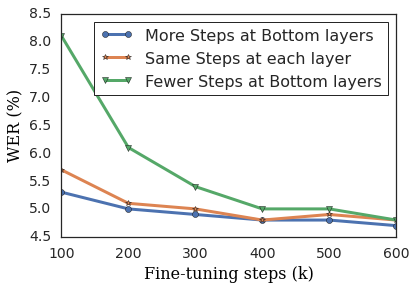}
         \caption{Figure 2b: Fine-tuning WERs for three schedules.}
          \label{fig:schedule}
     \end{subfigure}
\end{figure}

Firstly, we explore the effect of pre-training schedules and investigate how to distribute training steps on different encoder layers. As shown in Figure \ref{fig:pretrainsteps}, we tried three different schedules: more steps at bottom layers, same steps for all layers and fewer steps on bottom layers. Figure \ref{fig:schedule} shows that pre-training more steps at bottom layers, can converge faster in the fine-tuning stage. This is expected since the top layers cannot learn anything useful without having the well-trained bottom layers. 

\subsubsection{Different Number of Updated Layers at Each Step}
\label{sec:num_layers}
We update only $1$ layer at each step in Section \ref{sec:exp_lw_ssl}. Here, we vary the number of layers to be updated at each step. Table~\ref{tab:num_layers} shows that the proposed  algorithm is very robust to the number of layers being updated at each step. 

\subsubsection{Truncating Input Lengths}
\label{sec:input_length}
Furthermore, we reduce the training memory by truncating the input length of frames and fixing the updated layer per step to $1$. Memory is measured by running training in an Android emulator and it is only available if the updated layers per step is less than $5$. We have to estimate the training memory of the end-to-end algorithm  which updates 17 conformer layers at a time by extrapolating the data points using linear regression, because it exceeds the memory available. We find in Table \ref{tab:input_length} that the training memory of the proposed algorithm with input length $100$ ($439$MB) is $89.7\%$ smaller
 than the training memory of the end-to-end CPC algorithm ($7356$MB) without compromising performance.
\begin{table}[ht]
    \centering
    \caption{WER of Incremental layer-wise CPC algorithm with different number of bottom layers per step.}
    \label{tab:num_layers}
    \begin{tabular}{cccc}
        \toprule
       \textbf{$\#$ updated layers} & \multicolumn{2}{c}{\textbf{WER ($\%$)}} & \textbf{Memory}\\
        \textbf{per step}&  \textbf{MF} & \textbf{SF} & \textbf{Cost (MB)} \\
        \midrule
     End-to-end & {$4.5$} & {$6.1$}  & $7356$\\
     \midrule
         4 layers & {$4.5$} & {$6.1$} & $2007$\\
         2 layers & {$4.6$} & {$6.1$} & $1257$\\
         1 layer & {$4.6$} & {$6.1$} & $759$ \\
        \bottomrule
    \end{tabular}
\end{table}

\begin{table}[h]
    \centering
    \caption{WER of Incremental layer-wise CPC algorithm with truncated input lengths.}
    \label{tab:input_length}
    \begin{tabular}{ccccc}
        \toprule
       \textbf{Input length /} & \multicolumn{2}{c}{\textbf{WER ($\%$)}} & \multicolumn{2}{c}{\textbf{Memory Cost (MB)}} \\
        \textbf{Truncated $\%$}&  \textbf{MF} & \textbf{SF} & \textbf{Layer 1} &  \textbf{Layer 17} \\
        \midrule
     $686$ / 0$\%$& {$4.6$} & {$6.1$} &  $759$ &	$862$ \\
     \midrule
     $300$ / $56.3\%$ & {$4.6$} & {$6.1$} &  $410$ &	$590$ \\
         $200$ / $70.8\%$  & {$4.6$} & {$6.1$} & $313$ &	$515$   \\
          $100$ / $85.4\%$ & {$4.7$} & {$6.1$} & $246$ &	$439$\\
        \bottomrule
    \end{tabular}
\end{table}

\section{Conclusion}
\label{sec:conclusion}
In this paper, we propose an incremental layer-wise self-supervised learning algorithm for efficient speech domain adaptation on mobile devices. Extensive experimental results demonstrate that the proposed algorithm obtains a  Word Error Rate on the target domain $24.2\%$ better than supervised baseline and uses  $89.7\%$ less training memory than the end-to-end self-supervised learning algorithm. 


\bibliographystyle{IEEEbib}
\bibliography{refs}

\begin{thebibliography}{10}

\bibitem{he2019streaming}
Y.~He, T.~N. Sainath, R.~Prabhavalkar, I.~McGraw, R.~Alvarez, D.~Zhao,
  D.~Rybach, A.~Kannan, Y.~Wu, R.~Pang, et~al.,
\newblock ``Streaming end-to-end speech recognition for mobile devices,''
\newblock in {\em ICASSP 2019-2019 IEEE International Conference on Acoustics,
  Speech and Signal Processing (ICASSP)}. IEEE, 2019, pp. 6381--6385.

\bibitem{li2021better}
B.~Li, A.~Gulati, J.~Yu, T.~N. Sainath, C.-C. Chiu, A.~Narayanan, S.-Y. Chang,
  R.~Pang, Y.~He, J.~Qin, et~al.,
\newblock ``A better and faster end-to-end model for streaming asr,''
\newblock in {\em ICASSP 2021-2021 IEEE International Conference on Acoustics,
  Speech and Signal Processing (ICASSP)}. IEEE, 2021, pp. 5634--5638.

\bibitem{Bengio2012}
Y.~Bengio,
\newblock ``Deep learning of representations for unsupervised and transfer
  learning,''
\newblock in {\em Proceedings of ICML Workshop on Unsupervised and Transfer
  Learning}, 2012, pp. 17--36.

\bibitem{Sim2018}
K.~C. Sim, A.~Narayanan, A.~Misra, A.~Tripathi, G.~Pundak, T.~Sainath,
  P.~Haghani, B.~Li, and M.~Bacchiani,
\newblock ``Domain adaptation using factorized hidden layer for robust
  automatic speech recognition,''
\newblock in {\em Proc. Interspeech 2018}, 2018, pp. 892--896.

\bibitem{Ganin2016}
Y.~Ganin, E.~Ustinova, H.~Ajakan, P.~Germain, H.~Larochelle, F.~Laviolette,
  M.~Marchand, and V.~Lempitsky,
\newblock ``Domain-adversarial training of neural networks,''
\newblock {\em The Journal of Machine Learning Research}, vol. 17, no. 1, pp.
  2096--2030, 2016.

\bibitem{misra2021comparison}
A.~Misra, D.~Hwang, Z.~Huo, S.~Garg, N.~Siddhartha, A.~Narayanan, and K.~C.
  Sim,
\newblock ``A comparison of supervised and unsupervised pre-training of
  end-to-end models,''
\newblock 2021.

\bibitem{Chung2020_apc}
Y.~{Chung} and J.~{Glass},
\newblock ``Generative pre-training for speech with autoregressive predictive
  coding,''
\newblock in {\em ICASSP 2020 - 2020 IEEE International Conference on
  Acoustics, Speech and Signal Processing (ICASSP)}, 2020, pp. 3497--3501.

\bibitem{oord2018representation}
A.~v.~d. Oord, Y.~Li, and O.~Vinyals,
\newblock ``Representation learning with contrastive predictive coding,''
\newblock {\em arXiv preprint arXiv:1807.03748}, 2018.

\bibitem{baevski2020wav2vec}
A.~Baevski, H.~Zhou, A.~Mohamed, and M.~Auli,
\newblock ``wav2vec 2.0: A framework for self-supervised learning of speech
  representations,''
\newblock {\em arXiv preprint arXiv:2006.11477}, 2020.

\bibitem{mcmahan2017communication}
B.~McMahan, E.~Moore, D.~Ramage, S.~Hampson, and B.~A. y~Arcas,
\newblock ``Communication-efficient learning of deep networks from
  decentralized data,''
\newblock in {\em Artificial intelligence and statistics}. PMLR, 2017, pp.
  1273--1282.

\bibitem{kairouz2019advances}
P.~Kairouz, H.~B. McMahan, B.~Avent, A.~Bellet, M.~Bennis, A.~N. Bhagoji,
  K.~Bonawitz, Z.~Charles, G.~Cormode, R.~Cummings, et~al.,
\newblock ``Advances and open problems in federated learning,''
\newblock {\em arXiv preprint arXiv:1912.04977}, 2019.

\bibitem{ramaswamy2019federated}
S.~Ramaswamy, R.~Mathews, K.~Rao, and F.~Beaufays,
\newblock ``Federated learning for emoji prediction in a mobile keyboard,''
\newblock {\em arXiv preprint arXiv:1906.04329}, 2019.

\bibitem{hard2018federated}
A.~Hard, K.~Rao, R.~Mathews, S.~Ramaswamy, F.~Beaufays, S.~Augenstein,
  H.~Eichner, C.~Kiddon, and D.~Ramage,
\newblock ``Federated learning for mobile keyboard prediction,''
\newblock {\em arXiv preprint arXiv:1811.03604}, 2018.

\bibitem{guliani2021training}
D.~Guliani, F.~Beaufays, and G.~Motta,
\newblock ``Training speech recognition models with federated learning: A
  quality/cost framework,''
\newblock in {\em ICASSP 2021-2021 IEEE International Conference on Acoustics,
  Speech and Signal Processing (ICASSP)}. IEEE, 2021, pp. 3080--3084.

\bibitem{hinton2006fast}
G.~E. Hinton, S.~Osindero, and Y.-W. Teh,
\newblock ``A fast learning algorithm for deep belief nets,''
\newblock {\em Neural computation}, vol. 18, no. 7, pp. 1527--1554, 2006.

\bibitem{erhan2010does}
D.~Erhan, A.~Courville, Y.~Bengio, and P.~Vincent,
\newblock ``Why does unsupervised pre-training help deep learning?,''
\newblock in {\em Proceedings of the thirteenth international conference on
  artificial intelligence and statistics}. JMLR Workshop and Conference
  Proceedings, 2010, pp. 201--208.

\bibitem{huo2018decoupled}
Z.~Huo, B.~Gu, H.~Huang, et~al.,
\newblock ``Decoupled parallel backpropagation with convergence guarantee,''
\newblock in {\em International Conference on Machine Learning}. PMLR, 2018,
  pp. 2098--2106.

\bibitem{lowe2019putting}
S.~L{\"o}we, P.~O'Connor, and B.~S. Veeling,
\newblock ``Putting an end to end-to-end: Gradient-isolated learning of
  representations,''
\newblock {\em arXiv preprint arXiv:1905.11786}, 2019.

\bibitem{rudin1994total}
L.~I. Rudin and S.~Osher,
\newblock ``Total variation based image restoration with free local
  constraints,''
\newblock in {\em Proceedings of 1st International Conference on Image
  Processing}. IEEE, 1994, vol.~1, pp. 31--35.

\bibitem{gulati2020conformer}
A.~Gulati, J.~Qin, C.-C. Chiu, N.~Parmar, Y.~Zhang, J.~Yu, W.~Han, S.~Wang,
  Z.~Zhang, Y.~Wu, et~al.,
\newblock ``Conformer: Convolution-augmented transformer for speech
  recognition,''
\newblock {\em arXiv preprint arXiv:2005.08100}, 2020.

\bibitem{narayanan2019recognizing}
A.~Narayanan, R.~Prabhavalkar, C.-C. Chiu, D.~Rybach, T.~N. Sainath, and
  T.~Strohman,
\newblock ``Recognizing long-form speech using streaming end-to-end models,''
\newblock in {\em 2019 IEEE Automatic Speech Recognition and Understanding
  Workshop (ASRU)}. IEEE, 2019, pp. 920--927.

\end{thebibliography}

\end{document}